\documentclass[onecolumn]{pasj00}

\begin{document}
\SetRunningHead{Fujita et al.}{The Environment around RSGC~1 }

\title{The Environment around the Young Massive Star
Cluster RSGC~1 and HESS J1837--069}

\author{Yutaka \textsc{Fujita},\altaffilmark{1}
        Hiroyuki \textsc{Nakanishi},\altaffilmark{2}
	Erik \textsc{Muller},\altaffilmark{3}
	Naoto \textsc{Kobayashi},\altaffilmark{4}
	Masao \textsc{Saito},\altaffilmark{3,5}
	Chikako \textsc{Yasui},\altaffilmark{4}
	Hiroki \textsc{Kikuchi},\altaffilmark{1}
        and
        Keigo \textsc{Yoshinaga}\altaffilmark{1}
        }
\altaffiltext{1}{Department of Earth and Space Science, 
Graduate School of Science, Osaka University, 1-1 Machikaneyama-cho, 
Toyonaka, Osaka 560-0043}
\altaffiltext{2}{Department of Physics, Graduate School of Science 
and Engineering, Kagoshima University, 1-21-35 Korimoto, Kagoshima, 
Kagoshima 890-0065}
\altaffiltext{3}{National Astronomical Observatory of Japan, 
Osawa 2-21-1, Mitaka, Tokyo 181-8588}
\altaffiltext{4}{Institute of Astronomy, School of Science, 
University of Tokyo, 2-21-1 Osawa, Mitaka, Tokyo 181-0015}
\altaffiltext{5}{Joint ALMA Observatory (JAO), Alonso de Cordova 3107, 
Vitacura, Santiago, Chile} 


\maketitle

\begin{abstract}
We report on Mopra observations toward the young massive star cluster
 RSGC~1, adjoined by, and possibly associated with the gamma-ray source
 HESS J1837--069.  We measure the CO ($J=1$--0) distribution around the
 cluster and gamma-ray source, and find that the cluster is slightly
 higher than the velocity ranges associated with the Crux-Scutum arm. We
 reveal the cluster is associated with much less molecular gas compared
 with other young massive clusters in the Galaxy, Westerlund~1 (Wd~1)
 and 2 (Wd~2), which also radiate gamma-rays. We find no other
 structures that would otherwise indicate the action of supernova
 remnants, and due to the lack of material which may form gamma-rays by
 hadronic interaction, we conclude that the gamma-rays detected from
 HESS J1837--069 are not created through proton-proton interactions, and
 may more plausibly originate from the pulsar that was recently found
 near RSGC~1.
\end{abstract}

\section{Introduction}

Studies of red super-giants (RSGs) has been hampered by their small
numbers, in spite of their importance to investigations of stellar
evolution at stages immediately preceding supernova explosions. The
star cluster RSGC~1 is one of the rare clusters in the Galaxy containing
a significant number ($>10$) of RSGs \citep{fig06a,dav08a}. RSGC~1 is
fairly young, and very massive; with an initial mass estimated to be
$2\times 10^4$--$4\times 10^4\: M_\odot$, and an age of $\sim 10$~Myr
\citep{fig06a,dav08a}.

It has been shown RSGC~1 may host high-energy objects; the HESS
telescope array found the diffuse, $\sim 20$~pc gamma-ray source,
HESS~J1837--069 nearby to RSGC~1 \citep{aha05a}, implying cosmic-rays
(CRs) are accelerated around the cluster. CR acceleration in massive
young clusters is thought to arise through one of two processes: the CRs
are accelerated by pulsars in the clusters; or the CRs are accelerated
by shock waves formed by collisions of stellar winds or by supernova
explosions in the clusters. In the former, electrons are mainly
accelerated, while in the latter both electrons and protons can be
accelerated. In the case of RSGC~1, the former model has often been
supported. In fact, INTEGRAL found a hard X-ray source AX~J1838.0--0655
\citep{mol04a,bir04b,mal05a}, approximately co-located with RSGC~1,
where later observations using the Rossi X-Ray Timing Explorer (RXTE)
\citep{got08a} led to the discovery of 70.5~ms pulsations (see also
\cite{ana09a}). The pulsar PSR J1838--0655 is a rotation-powered pulsar
with spin-down luminosity $\dot{E}=5.5\times 10^{36}\rm\: erg\: s^{-1}$
and characteristic age $\tau_c=2.3\times 10^4$~yr: the properties of the
pulsar and the pulsar wind nebular (PWN) have also been theoretically
studied by \citet{mat09b} and \citet{lin09a}. The hydrogen column
density toward the pulsar is almost the same as those for stars in
RSGC~1, which may imply that the pulsar is associated with the cluster
\citep{got08a}. The discovery of the pulsar supports the scenario where
the observed gamma-rays are formed by inverse Compton scattering of the
CR electrons in the PWN (leptonic process; \cite{ato96a,tan10a}).

RSGC~1 is not the only young massive cluster in the Galaxy that radiates
gamma-rays: other canonical examples being Westerlund~1 (Wd~1) and~2
(Wd~2).  Gamma-rays are observed from a wide ($\sim 160$~pc) region
around Wd~1 \citep{abr12a}, although the source of the CRs that emit the
gamma-rays has not been identified. In contrast to RSGC~1, the extremely
large spatial scale would prevent CR electrons prevailing in the
gamma-ray emission region before they are affected by cooling
\citep{abr12a}. Thus, in the case of Wd~1, CR protons may be responsible
for generating the observed gamma-rays, through their interaction with
gas protons and the pion-decay process (hadronic process), and in fact,
Wd~1 has sufficient gas mass to enable this process \citep{abr12a}.  In
the case of Wd~2, the spatial size of the gamma-ray emission region is
$\sim 6$--31~pc (depending on the radial distance;
\cite{aha07a,abr11a}). Since pulsars have been detected in the gamma-ray
emission region \citep{abd09e}, PWNe are likely responsible for
generating the observed gamma-rays. However, as CR protons can certainly
be accelerated in stellar winds or at the shock waves of supernova
remnants (SNRs), gamma-ray formation through the hadronic process cannot
be ruled out \citep{fuj09d,abr11a}. Nearby the young cluster; the Arches
cluster, at the Galactic center, is another gamma-ray source, 3EG
J1746--2851, which may radiate via inverse Compton scattering of the
radiation field of the cluster \citep{yus03a}. However, since the
Galactic center is well populated with a wide variety of putative such
objects, and is somewhat dynamic and confused, it may be premature to
conclude definitively that this is the precise mechanism operating in
this case \citep{poh05a,tat12a}.

In this paper, we present the results of our observations of molecular
material towards RSGC~1, traced by $^{12}$CO(1-0) at 115.271 GHz. This
study is motivated by the discovery of the pulsar near the cluster,
which implies the gamma-rays need not necessarily have a leptonic
origin: The existence of the pulsar is significant in suggesting that CR
protons could have been accelerated at the SNR associated with the
pulsar. If sufficient gas exists around the cluster, the protons
contained in the gas may collide with the CR protons, generating and
radiating gamma-rays. The study of the gas distribution around the
cluster may also give us clues regarding the evolution of massive
clusters, bearing in mind the age of RSGC~1 is 10--14~Myr
\citep{dav08a}, and those of Wd~1 and 2 are $\sim 5$~Myr and $\sim
2$~Myr, respectively \citep{lim13a,car13b} and that difference of the
ages necessarily implies the gas distribution around those clusters will
be at different evolutionary stages, and therefore, will be dispersed
differently.

CR acceleration in massive young clusters is also important in the
context of high-energy CR protons and their role in the so-called 'knee'
in the the CR energy spectrum ($\sim 10^{15}$~eV). Usually thought to be
accelerated in the shock waves of isolated SNRs, the observed energy of
CRs around such SNRs is generally much smaller than $10^{15}$~eV
\citep{but09a}. In a massive cluster however, overlapping SNRs would
develop into super-bubbles with shocks whose larger size and longer
lifetime may accelerate CRs up to the knee, and thereby generating TeV
gamma-ray emissions \citep{but09a}. If this is the case, and if the gas
density around the cluster is high, strong and wide-spread gamma-rays of
hadronic origin would be observed from the cluster.

The radial velocity of RSGC~1 is well constrained thorough the
observations of stars in the cluster. In this study, we assume that the
radial velocity is $V_{\rm LSR}=123\pm 4\rm\: km\: s^{-1}$ and the
distance to the cluster is $d=6.60\pm 0.89$~kpc \citep{dav08a}. Thus,
$1'$ corresponds to 1.9~pc. The position of the field center corresponds
to the cluster center at R.A.=\timeform{18h37m58s},
decl.=\timeform{-6D52'53''.0} (J2000.0), as shown in \citet{fig06a}.

\section{Observations}

Observations of the $J=1$--0 transition of $^{12}$CO were made with
Mopra\footnote{Operation of the Mopra radio telescope is made possible
by funding from the National Astronomical Observatory of Japan, the
University of New South Wales, the University of Adelaide, and the
Commonwealth of Australia through CSIRO} 22~m single-dish radio
telescope in Australia, in September 2012.  The half-power beam width of
the telescope was $33''$ (at the $^{12}$CO(1-0) transition frequency of
115~GHz), sampled with a spectrometer over 4096 channels. The resulting
velocity resolution is $0.087\rm\: km\: s^{-1}$; pixel size is $15''$
and the rms noise per channel is $\sim 0.4$~K (efficiency not
corrected). We compute the main-beam temperatures using an efficiency of
0.55, as recommended for extended structures by \citet{lad05a}. This
efficiency value was confirmed to be consistent by our concurrent
observations of M17SW, which yielded peak antenna temperatures of
25$\pm1$ K, and is generally consistent with archived SEST
results\footnote{http://www.apex-telescope.org/sest/html/telescope-calibration/calib-sources/m17sw.html}
where the {\it extended} beam efficiency is assumed to be 20\% higher
than the cited {\it compact} (as it is for Mopra) of 0.45. The data were
reduced using the Australia Telescope National Facility analysis
programs, {\sc livedata} and {\sc
gridzilla}\footnote{http://www.atnf.csiro.au/computing/software/livedata/},
and they were analyzed using AIPS software
package\footnote{http://www.aips.nrao.edu/index.shtml}.

\section{Results}

\begin{figure}
  \begin{center}
    \FigureFile(80mm,80mm){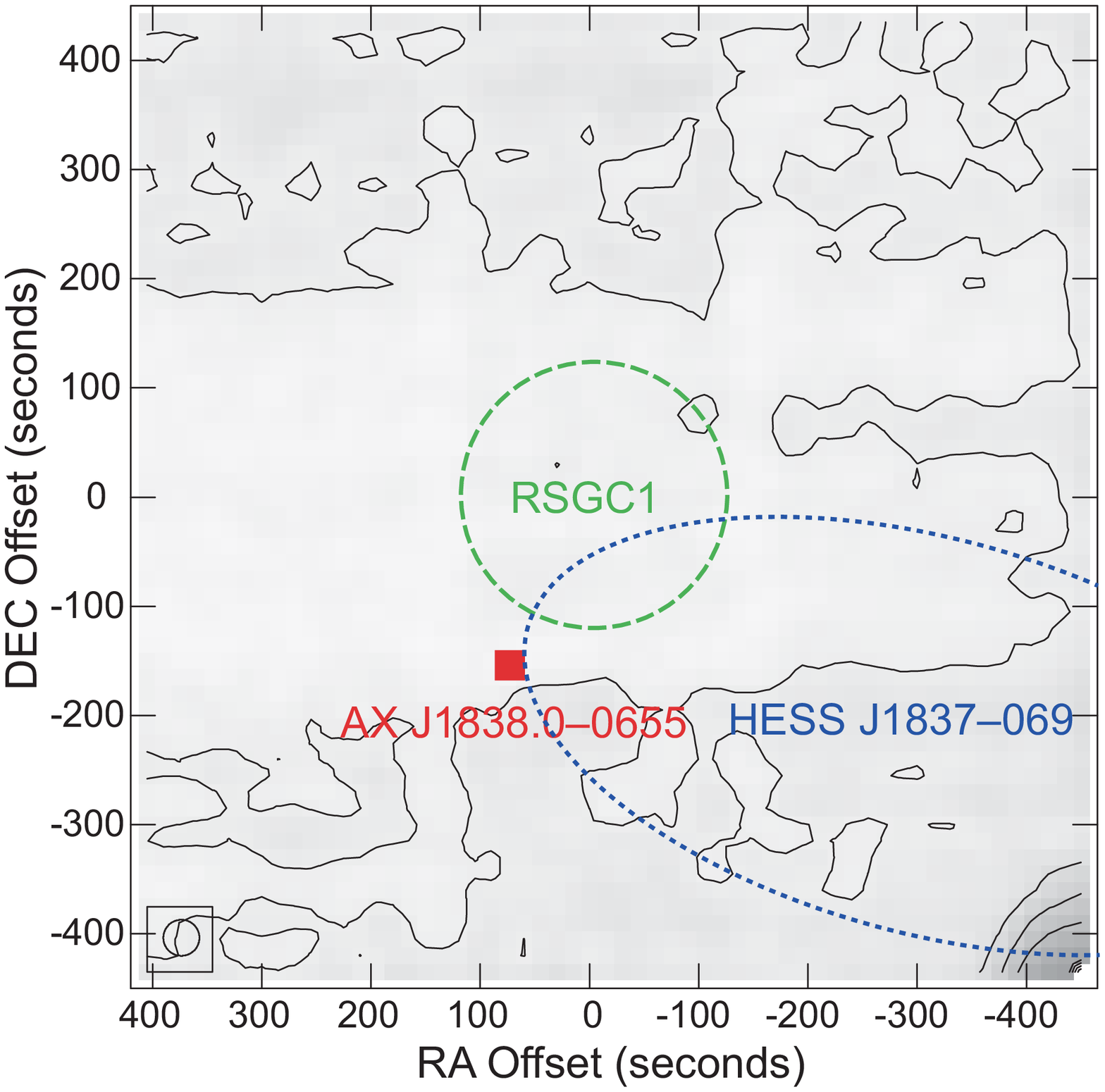}
  \end{center}
  \caption{Integrated intensity map of the $^{12}$CO emission for
 $116\leq V_{\rm LSR}\leq 130\rm\: km\: s^{-1}$ with contour levels of
 3, 6, 9, and $12\rm\: K\: km\: s^{-1}$. The origin of the coordinate is
 R.A.=\timeform{18h37m58s}, and decl.=\timeform{-6D52'53''.0} (J2000.0).  The
 dashed circle is the approximate extent of the massive star cluster
 RSGC~1 \citep{fig06a}. The square shows the position of
 AX~J1838.0--0655. The dotted ellipse represents the observed excess of
 the TeV emission identified as HESS~J1837--069
 \citep{aha06c}.}\label{fig:image}
\end{figure}

An integrated intensity map of the observed $15'\times 15'$ region
centered on RSGC~1 is shown in figure~\ref{fig:image}. These data are
integrated across the velocity range corresponding to the that of
RSGC~1; $116\leq V_{\rm LSR}\leq 130\rm\: km\: s^{-1}$. The gamma-ray
emission region appears to cover a large portion of the field
\citep{aha06c}, and the X-ray source AX~J1838.0--0655 is shown, located
to the southeast of RSGC~1. Significantly, we do not find any obvious
unusually high column densities of CO at this velocity range
corresponding to RSGC~1.

\begin{figure}
  \begin{center}
    \FigureFile(80mm,80mm){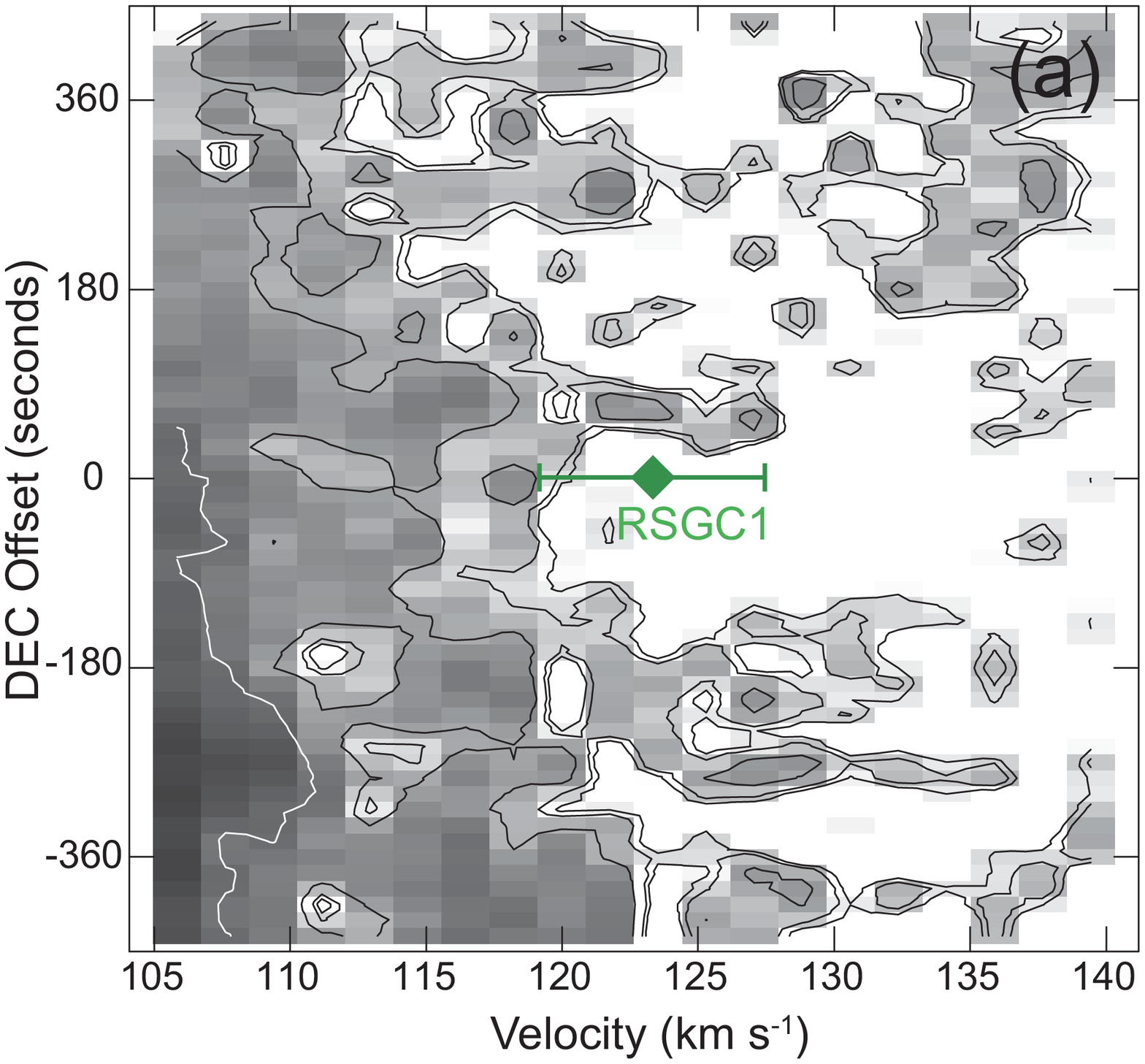} \FigureFile(80mm,80mm){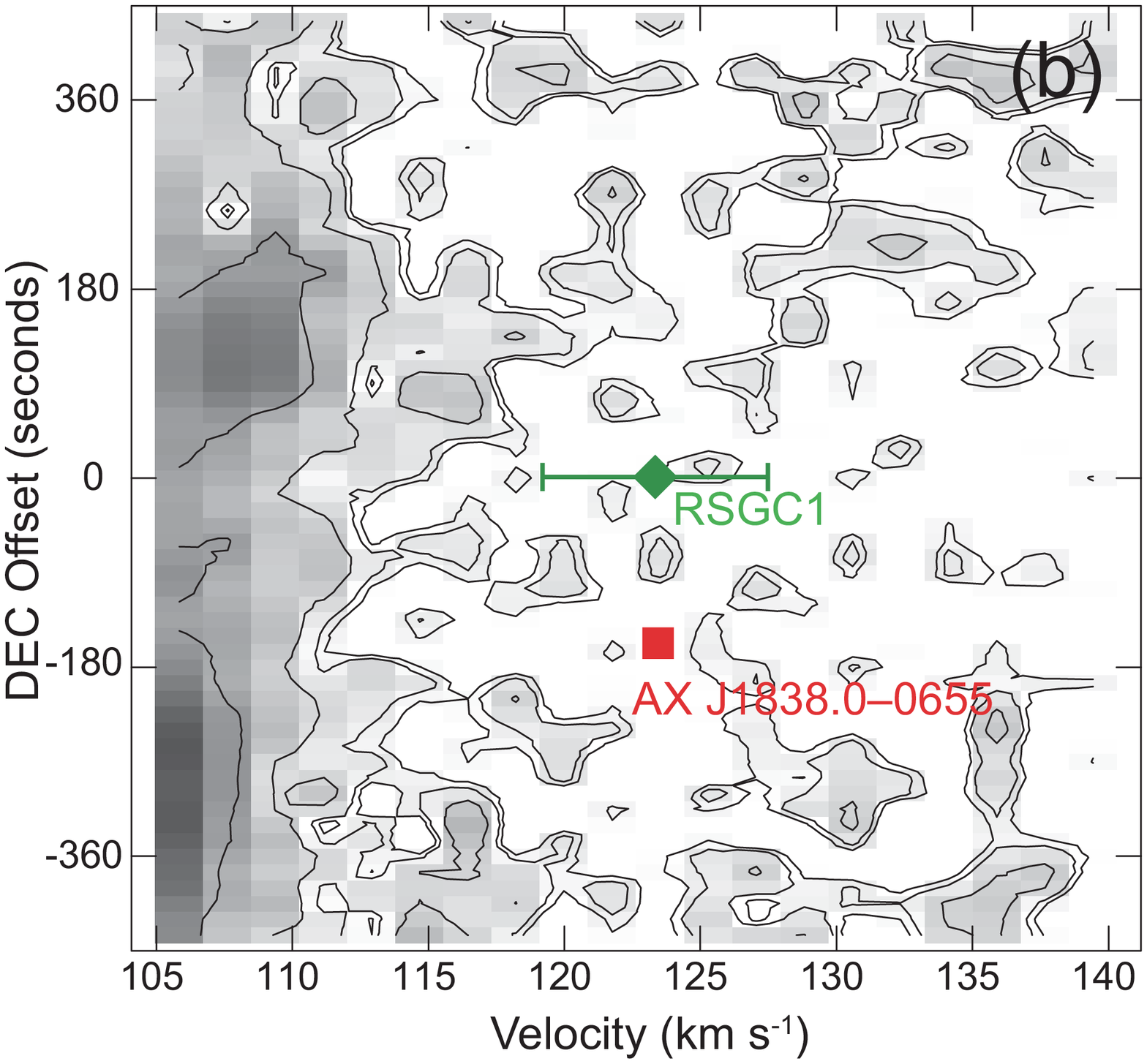}
  \end{center}
  \caption{(a) Velocity versus equatorial latitude diagram for $^{12}$CO
 $(J = 1-0)$ emission integrated over a RA offset range of $-400''$ to
 $-50''$ (see figure~\ref{fig:image}). Contour levels are 0.012, 0.04,
 0.12, and 0.4~K. (b) Same as (a) but it is integrated over a RA offset
 range of $-100''$ to $100''$ (see figure~\ref{fig:image}). Contour
 levels are 0.02, 0.067, 0.2, and 0.67~K. The ratio of contour levels
 between (a) and (b) is that of the integration length in the RA
 direction. The origin of the vertical axis is the position of RSGC~1
 (decl.=\timeform{-6D52'53''.0}; J2000.0).  Positions of RSGC1 and
 AX~J1838.0--0655 are shown by the diamond and the square,
 respectively. We assume that the velocity of AX~J1838.0--0655 is the
 same as that of RSGC~1.}\label{fig:PV}
\end{figure}

\begin{figure}
  \begin{center}
    \FigureFile(160mm,160mm){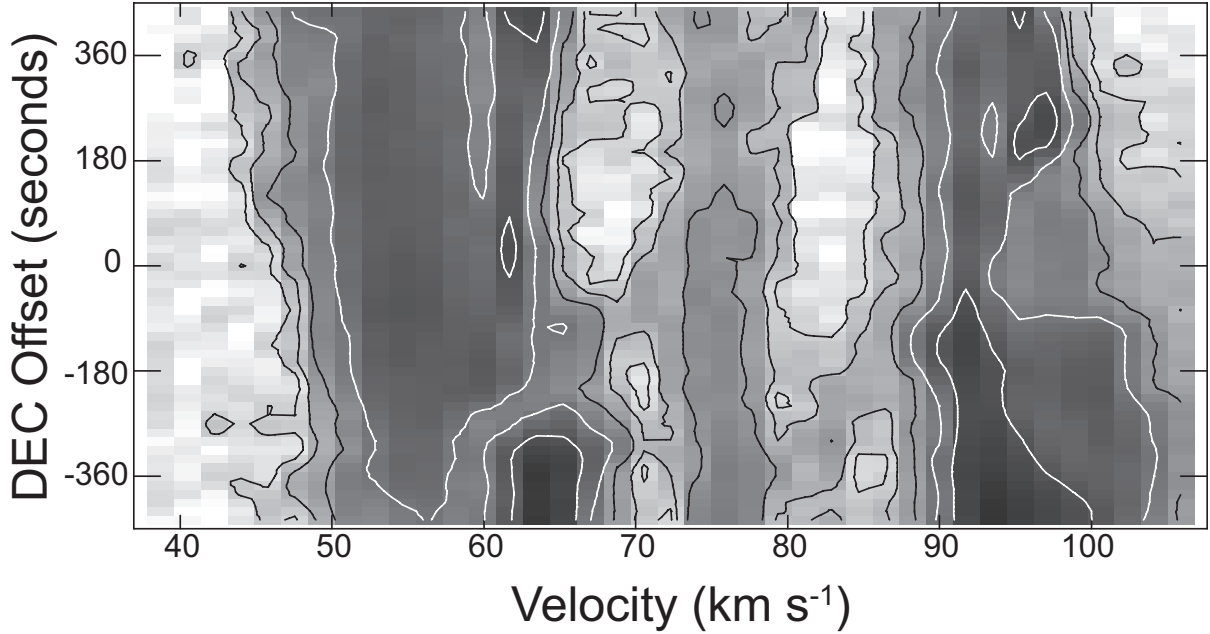}
  \end{center}
  \caption{Same as figure~\ref{fig:PV}a but for $38\leq V_{\rm LSR} \leq
106\rm \: km\: s^{-1}$. Contour levels are 0.2, 0.4, 0.8, 1.6, and
3.2~K.}\label{fig:PV_f}
\end{figure}

Figure~\ref{fig:PV}a shows CO ($J=1$--0) velocity-latitude diagram
integrated over a RA offset range of $-400''$ to $-50''$ (see
figure~\ref{fig:image}), where the velocity channels are binned over 20
channels. This RA range contains the gamma-ray radiating region (dotted
line in figure~\ref{fig:image}).  While the gas is deficient for $V_{\rm
LSR}\gtrsim 120\rm\: km\: s^{-1}$, it is abundant for $V_{\rm
LSR}\lesssim 120\rm\: km\: s^{-1}$. The gas rich region at $V_{\rm
LSR}\lesssim 120\rm\: km\: s^{-1}$ apparently corresponds to the
Crux-Scutum arm that is located at $V_{\rm LSR}\sim 95\rm\: km\: s^{-1}$
(see also figure~\ref{fig:PV_f}).  RSGC~1 has a velocity slightly higher
than the Crux-Scutum arm. There seems to be no structure associated with
the gamma-ray radiating region (from $-400''$ to 0 in the DEC offset;
see figure~\ref{fig:image}). Figure~\ref{fig:PV}b is the same as
figure~\ref{fig:PV}a but it is integrated over a RA offset range of
$-100''$ to $100''$ (see figure~\ref{fig:image}) in order to study the
region including RSGC~1 and AX~J1838.0--0655. We assume in the figure
that the velocity of AX~J1838.0--0655 is the same as that of RSGC~1. CO
emission around RSGC~1 and AX~J1838.0--0655 is rather small. If we adopt
a conversion factor of $1.8\times 10^{20}\rm cm^{-2}\rm (K\: km\:
s^{-1})^{-1}$ \citep{dam01a}, the column density in the region for
$116\leq V_{\rm LSR}\leq 130\rm\: km\: s^{-1}$ is $N_{\rm H_2}\lesssim
1\times 10^{21}\rm\: cm^{-2}$. However, it is not clear whether the lack
of the emission is physically associated with RSGC~1 and
AX~J1838.0--0655.

\section{Discussion}

Figure~\ref{fig:PV} does not reveal any notable nor peculiar structure
in the distribution of the molecular component around the velocity of
RSGC~1 ($V_{\rm LSR}\sim 123\rm\: km\: s^{-1}$).  We may constrain the
mass of molecular gas associated with RSGC~1 by estimating the mass in a
velocity range of $116\leq V_{\rm LSR}\leq 130\rm\: km\: s^{-1}$. We
chose this velocity range because the range is not much affected by the
Crux-Scutum arm.  Note that since RSGC~1 is located on the peak of the
Galactic rotation curve in that direction (figure~4 in \cite{dav08a}),
molecular gas with $V_{\rm LSR}\gtrsim 123\rm\: km\: s^{-1}$ does not
rotate with the Galactic disk. For the conversion factor of $1.8\times
10^{20}\rm cm^{-2} (K\: km\: s^{-1})^{-1}$ \citep{dam01a}, the molecular
hydrogen mass in the entire $15'\times 15'$ field for the velocity range
is $M_g=1.2\times 10^{4}\: M_\odot$. The mass of the gas that is
gravitationally associated with the cluster could be even smaller,
because the apparent size of the cluster is smaller than the $15'\times
15'$ field (figure~\ref{fig:image}). We also derive the gas mass in the
gamma-ray radiating field inside the dotted line in
figure~\ref{fig:image} for $116\leq V_{\rm LSR}\leq 130\rm\: km\:
s^{-1}$. We found that it is only $M_{g\gamma}=2.5\times 10^3\:
M_\odot$.

The rarity of the gas around RSGC~1 is notable after comparison of the
gas mass with other young massive star clusters such as Wd~1 and 2.
\citet{abr12a} estimated that the hydrogen molecular density within a
radius of $1.1^\circ$ from Wd~1 is $\sim 12\rm\: cm^{-3}$ (see also
\cite{dam01a,yam03a}). If the hydrogen is spherically distributed, the
total mass would be $1.4\times 10^{6}\: M_\odot$. In the case of Wd~2,
two molecular clouds are associated with the cluster and their total
mass is $1.7\times 10^5\: M_\odot$ \citep{fur09a}. Here we can note the
difference of the gas mass may be due to that of cluster ages; evolved
massive stars and/or supernovae in RSGC~1 may have injected sufficient
energy to disperse or even photo-disassociate the molecular gas during
the cluster lifetime.

We examine the foreground (i.e. lower velocity regime) of RSGC~1.
Figure~\ref{fig:PV_f} shows that molecular material is concentrated at
$V_{\rm LSR}\sim 55$ and~$95\rm\: km\: s^{-1}$. In the direction of
RSGC~1 ($l=25.3^\circ$ and $b=-0.2^\circ$), these concentrations are the
Carina-Sagittarius arm and the Crux-Scutum arm, respectively (figure~3
in \cite{mom06a}). In the direction of the gamma-ray radiating region
(from $-400''$ to 0 in the DEC offset; see figure~\ref{fig:image}), we
find no obvious structures that may indicate the existence of SNRs
around the velocities of the arms where SNRs are often found. Thus, we
consider it unlikely that the gamma-rays observed in the direction of
RSGC~1 are emitted by foreground SNRs.

If the gamma-rays observed around RSGC~1 have a hadronic origin, they
should be created through interactions between CR protons and the
protons in the gas around the cluster. However, the deficiency of the
gas around the cluster strongly suggests that there is a lack of the
target protons for the interactions, without which the observed
gamma-ray emission would require extremely large flux of CR protons to
produce a given gamma-ray luminosity. Based on the model by
\citet{fuj09c}, we estimate the total energy of the CR protons required
to produce the observed gamma-ray luminosity for the measured molecular
gas-mass ($M_{g\gamma}=2.5\times 10^3\: M_\odot$) as $E_{\rm CR}\sim
10^{52}$~erg.  This energy is unreasonably large, and corresponds to the
energy of CRs accelerated by $\sim 100$ supernovae (assuming an
acceleration efficiency of $\sim$10\%). Even if $\sim 100$ supernovae
have exploded, CRs contained in older SNRs must have been expelled by
newer SNRs. That is, if the total gas and CR energy of $10^{53}$~erg is
confined in the gamma-ray emitting volume ($\sim 10^4\rm\: pc^3$), the
pressure there should be $\sim 3\times 10^{-7}\rm\: dyn\: cm^{-2}$.
This is much larger than the typical pressure in the Galactic disk
($\sim 10^{-12}\rm\: dyn\: cm^{-2}$), and the gas and CRs in that volume
cannot be physically be confined.  Therefore, we can confidently reject
the hadronic scenario for gamma-ray production towards RSGC~1, in
contrast with Wd~1 and~2. Although our field does not include the whole
gamma-ray radiating region (figure~\ref{fig:image}), it contains most of
it, and thus the conclusion is solid. A lack of spatial correlation
between the gamma-ray radiating field and the CO emission at the
right-bottom corner in figure~\ref{fig:image} also supports the
conclusion. The recent discovery of the pulsar PSR~J1838-0655 at the
position of AX~J1838.0--0655 suggests that the PWN associated with it is
the gamma-ray source \citep{got08a}, and further supports our findings
here.

\section{Conclusions}

We observed the distribution of $^{12}$CO ($J=1$--0) emission towards
and around the massive young star cluster RSGC~1 with Mopra 22~m radio
telescope. The cluster is located just outside the Crux-Scutum arm and
we find the gas mass around the cluster is much smaller than those of
other massive young clusters in the Galaxy, such as Wd~1 and~2. The gas
distribution of the foreground region is rather smooth and there seem to
be no evidence of active activities that might imply additional
energetic processes. The low gas mass around the cluster indicates that
the gamma-ray emission around the cluster (HESS~J1837--069) is
insufficient for providing enough mass to support the hadronic scenario
for gamma-ray production, and we suggest the CR emission appears to be
coming from the pulsar found near the cluster.

\bigskip

This work was supported by KAKENHI (YF: 23540308).


\end{document}